\documentclass[%
 reprint,
superscriptaddress,
 amsmath,amssymb,
 aps,
prl,
notitlepage
]{revtex4-1}
\usepackage{color}

\usepackage{graphicx}
\usepackage{dcolumn}
\usepackage{bm}
\usepackage{color}

\usepackage[dvipsnames]{xcolor}
\usepackage[utf8]{inputenc}
\usepackage{wrapfig}
\usepackage[]{graphicx,xcolor}
\usepackage{tabularx}
 \usepackage{booktabs}
\usepackage{textcomp}
\usepackage{amsmath}
\usepackage{amssymb}
\usepackage[]{graphics}
\usepackage{amssymb}
\usepackage{amsmath}
\usepackage{color}
\usepackage{ifpdf}
\usepackage{lipsum}
\usepackage{siunitx}
\usepackage{braket}
\usepackage{soul}
\usepackage{tikz}
\usepackage{color}
\usepackage[dvipsnames]{xcolor}
\usepackage[breaklinks,
            colorlinks,
            urlcolor=blue,
            linkcolor=blue,
            anchorcolor=blue,
            citecolor=blue]{hyperref}

\usepackage{appendix}
\usetikzlibrary{automata,positioning}

\begin{document}

\title{Light-cone-proximal quasi-BICs for chiral lasing at grazing angles}

\author{Dmitrii Gromyko}
\affiliation{Science, Mathematics, and Technology (SMT), Singapore University of Technology and Design (SUTD), 8 Somapah Road, Singapore 487372.}
\affiliation{Department of Electrical and Computer Engineering, National University of Singapore, 4 Engineering Drive 3, Singapore 117583.}

\author{Cheng-Wei Qiu}
\email{chengwei.qiu@nus.edu.sg}
\affiliation{Department of Electrical and Computer Engineering, National University of Singapore, 4 Engineering Drive 3, Singapore 117583.}

\author{Kirill Koshelev}
\email{kirill.koshelev@anu.edu.au}
\affiliation{Department of Electronic Materials Engineering, Research School of Physics,
Australian National University, Canberra, ACT, Australia}

\author{Lin Wu}
\email{lin\_wu@sutd.edu.sg}
\affiliation{Science, Mathematics, and Technology (SMT), Singapore University of Technology and Design (SUTD), 8 Somapah Road, Singapore 487372.}

\begin{abstract}
Chiral quasi-bound states in the continuum (q-BICs) have recently emerged in metaphotonics as resonances that combine ultrahigh quality factors with near-unity circular polarization. However, these states are typically confined to the $\Gamma$-point (normal incidence) due to their symmetry-protected origins. 
We propose a new mechanism for realizing light-cone-proximal chiral q-BICs at large oblique angles, enabled by the divergence of the radiative local density of states near the light cone. Using dielectric metasurfaces with a monoclinic lattice and broken in-plane mirror symmetry, we demonstrate that tuning the lattice angle allows for robust control of these resonances.
The resulting chiral q-BICs exhibit near-unity circular dichroism in transmission and fully circularly polarized emission at angles exceeding $50^\circ$ from normal. This approach paves the way for directional chiral lasing at grazing angles and for photonic devices operating efficiently in off-normal geometries.
\end{abstract}

\maketitle

\textcolor{blue}{\it Introduction}.--- Dielectric metaphotonics has emerged as a powerful platform for efficient nanoscale light manipulation~\cite{kuznetsov2016optically,tseng2020dielectric}. Over the past decade, metastructures supporting optical resonances have enabled enhanced light-matter interactions~\cite{koshelev2020dielectric,feng2023dual,ma2025quantum}, driven by resonances with exceptionally high quality factors ($Q$-factors) \cite{huang2023ultrahigh}. 
A key focus has been on photonic bound states in the continuum (BICs) \cite{hsu2016bound,kang2023applications}, which suppress radiative losses either through symmetry protection (symmetry-protected BICs) or destructive interference (accidental BICs) \cite{koshelev2023bound}. These states are characterized by polarization vortices in the far field~\cite{zhen2014topological}.
Their symmetry-broken counterparts, quasi-BICs (q-BICs), feature finite but high $Q$-factors~\cite{koshelev2018asymmetric,liu2022thermal}, and arise naturally from fabrication imperfections~\cite{kuhne2021fabrication}. These modes have been widely employed in biosensing~\cite{tittl2018imaging}, nonlinear optics~\cite{liu2019high,zograf2022high,gromyko2025enabling}, and lasing ~\cite{kodigala2017lasing,huang2020ultrafast,wu2020room,wang2020generating,hwang2021ultralow,do2025room,cui2025ultracompact}.

Introducing chirality into metaphotonic structures offers an additional degree of freedom to control light-matter interactions~\cite{hentschel2017chiral,schaferling2017chiral,koshelev2023nonlinear}. In particular, generating circularly polarized and collimated emission is highly desirable, as most conventional light sources emit linearly polarized or unpolarized light with poor directionality~\cite{curto2010unidirectional,liu2025single}. The recently proposed concept of chiral q-BICs addresses this challenge by combining high $Q$ resonances with strong circular polarization in the far field~\cite{gorkunov2020metasurfaces, overvig2021chiral,chen2023observation,kuhner2023unlocking,toftul2024chiral,sinev2025chirality}. However, most demonstrations of chiral q-BIC lasing remain limited to normal-incidence emission, as the underlying q-BICs typically originate from symmetry-protected BICs at the $\Gamma$-point of the Brillouin zone~\cite{zhang2022chiral,maksimov2022circularly,gromyko2024unidirectional,deng2025chiral,chen2025ultraviolet}.

Chiral lasing at oblique angles with high directionality has received comparatively little attention, though it holds significant potential~\cite{ha2018directional,ha2022room}. Existing approaches remain complex: synthetic valley engineering enables large-angle emission~\cite{chen2023compact,xing2025chiral}, but cannot suppress undesired output at normal incidence. Brillouin zone folding can generate chirality points ($C$-points)~\cite{jeong2025obtuse}, yet hampers selective amplification due to uniformly high $Q$-factors across momentum space. A simpler, more robust route to directional chiral lasing remains largely unexplored.

\begin{figure}[h!]
\centering
\includegraphics[width=1\linewidth]{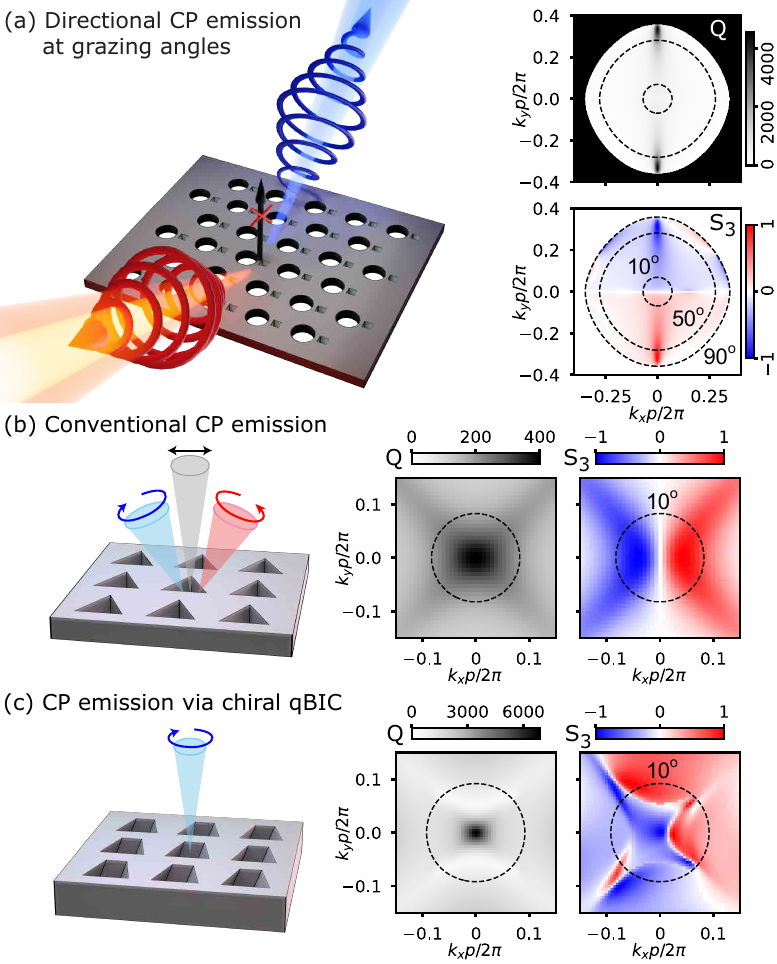}
\caption{ 
Mechanisms of circularly polarized (CP) emission from resonant metasurfaces. 
Left: schematic illustrations. 
Right: mode 
$Q$-factor and circular polarization degree 
$S_3$ in momentum space.
(a) Grazing-angle chiral emission via light-cone-proximal q-BICs (this work).
(b) Conventional 
$C$-point-based approach (adapted from~\cite{liu2019circularly}).
(c) Normal-incidence chiral emission via symmetry-protected q-BIC (adapted from~\cite{chen2023observation}).
}
\label{fig:1}
\end{figure}

In this Letter, we introduce a double symmetry-breaking strategy for realizing chiral q-BICs in resonant dielectric metasurfaces at large oblique angles approaching grazing incidence, as illustrated in Fig.~\ref{fig:1}(a). Unlike conventional approaches [Figs.~\ref{fig:1}(b,c)], we employ accidental BICs, which, upon breaking the lattice symmetry, split into q-BICs and $C$-points.
Near the light cone, lattice deformation shifts the q-BICs and $C$-points differently in momentum space due to the rapid increase in the radiative density of states. This enables robust alignment of the $Q$-factor peak with a $C$-point at oblique incidence, forming what we term a ``light-cone-proximal chiral q-BIC''. Full-wave simulations confirm that such resonances can generate highly directional, circularly polarized (CP) emission at angles exceeding $50^\circ$ from normal. This opens a practical route toward compact, efficient sources of circularly polarized light with engineered emission directionality.

\begin{figure*}
\centering
\includegraphics[width=1\linewidth]{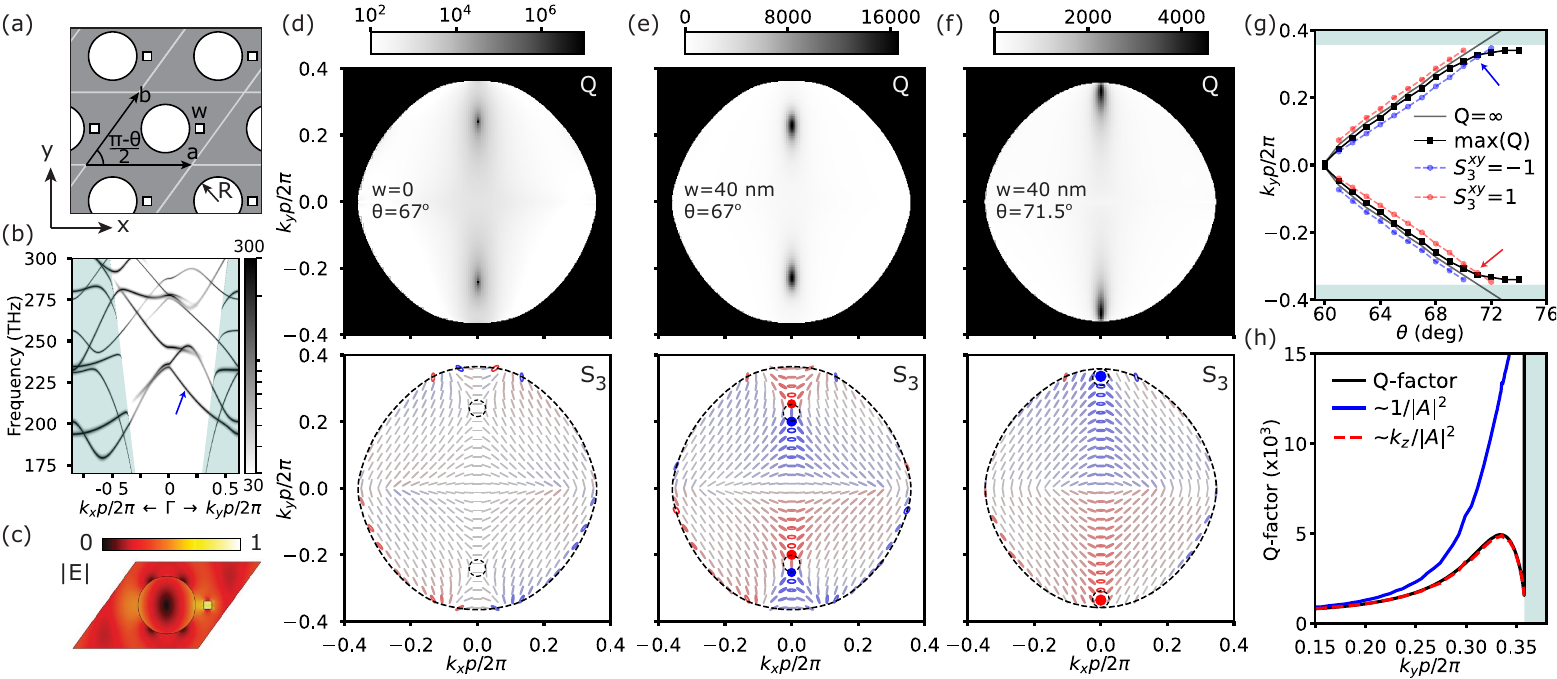}
\caption{ 
Framework of light-cone-proximal chiral q-BICs. 
(a) Unit cell schematic.
(b) Scattering matrix norm showing band structure; blue arrow marks the target mode. The green shading in panels (b,g,h) indicates the area in $\mathbf{k}-\omega$ space under the light cone.
(c) Electric field norm of the mode.
(d-h) Evolution of BICs into q-BICs and 
$C$-points via lattice angle $\theta$ and square hole width $w$:
(d) Accidental BICs at 
$\theta=67^\circ$,
$w=0$.
(e) Splitting into q-BICs and $C$-points at 
$w=40$~nm.
Dashed circles indicate the local maxima of the $Q$-factor. 
(f) Merging of q-BIC and 
$C$-point via further 
$\theta$-perturbation.
(g) Trajectories of accidental BICs (gray), q-BICs (black), and 
$C$-points (blue/red) in parameter space.
(h) Mode $Q$-factor evolution (black) compared to analytic model in Eq.~\eqref{gamma_rad_final} (red dashed), contributed by radiative coupling $1/|A|^2$ (blue) and inverse vacuum LDOS. 
}
\label{fig:2}
\end{figure*}

\textcolor{blue}{\it Concept}.---  
To demonstrate the concept of light-cone-proximal BICs, we consider a resonant metasurface comprising a high-index dielectric membrane with a monoclinic unit cell containing a larger circular and a smaller rectangular hole [Fig.~\ref{fig:1}(a)]. The fundamental optical mode exhibits two sharp $Q$-factor peaks in momentum space near the light cone, which coincide with $C$-points—locations where the far-field degree of circular polarization (DCP), defined as the third Stokes parameter $S_3 = (I^R - I^L)/(I^R + I^L)$, reaches $\pm1$, with $I^{R,L}$ denoting right- and left-circularly polarized radiation intensities. This alignment of $Q$-factor maxima and $C$-points results in highly directional, circularly polarized emission in two beams at angles exceeding $50^\circ$. Enhancement of emission along other directions is much less and can be neglected. 

In contrast, conventional $C$-point designs~\cite{liu2019circularly,wang2020routing,seo2021circularly,tian2022optical} do not align $Q$-factor maxima with $C$-points [Fig.~\ref{fig:1}(b)]; the $Q$-factor peak at the $\Gamma$-point corresponds to linearly polarized emission, while $C$-points occur at lower $Q$-factors within narrow angular ranges ($<10^\circ$). A more recent alternative, chiral q-BICs based on symmetry-protected BICs, produce circularly polarized emission confined to normal incidence, as the $Q$-factor rapidly drops away from the $\Gamma$-point [Fig.~\ref{fig:1}(c)].

\textcolor{blue}{\it Metasurface design \& theoretical framework}.--- 
We consider a resonant metasurface composed of a high-index dielectric membrane (thickness $h = 200$ nm, permittivity $\varepsilon = 12.25$) suspended in air and patterned with a monoclinic unit cell [Fig.~\ref{fig:2}(a)]. The lattice is defined by vectors $\mathbf{a} = 2p\sin(\theta/2)\hat{x}$ and $\mathbf{b} = p\sin(\theta/2)\hat{x} + p\cos(\theta/2)\hat{y}$, where $\theta$ is the lattice angle and $p = 520$ nm. Each unit cell contains a circular hole of radius $R = 140$ nm and a smaller square hole of width $w = 40$ nm placed 200 nm away from the circular hole center. The impact of a substrate is discussed in SI \cite{supp}.

We simulate the optical response using eigenmode and scattering calculations via the open-source RCWA package Inkstone \cite{alexysong_inkstone}, which is well-suited for periodic structures near light cones and Rayleigh anomalies \cite{tikhodeev2002quasiguided,gromyko2022resonant}. Details of the numerical methods and comparisons with COMSOL finite-element simulations are provided in SI~\cite{supp}. The dispersion of the mode of interest is marked by an arrow in Fig.~\ref{fig:2}(b), and its electric field profile is shown in Fig.~\ref{fig:2}(c).

In the symmetric case ($w=0$), a hexagonal lattice ($\theta = 60^\circ$) supports a symmetry-protected BIC with topological charge $-2$ at the $\Gamma$-point, which splits into two accidental BICs as $\theta$ increases~\cite{yoda2020generation,wang2022realizing}. For a monoclinic lattice with $\theta = 67^\circ$ and broken six-fold symmetry, the $Q$-factor map in momentum space [Fig.~\ref{fig:2}(d)] reveals two divergent peaks at $k_x = 0$, $k_yp/2\pi \approx \pm 0.25$, each associated with a polarization vortex (topological charge $-1$) visible in the $S_3$ map.

To induce chirality and break the symmetry for the second time, we introduce the square hole ($w = 40$ nm), which breaks both $zy$-mirror and two-fold rotational symmetry [Fig.~\ref{fig:2}(e)]. Each accidental BIC then splits into a pair of $C$-points \cite{zhen2014topological} and a q-BIC, which is defined as the position of the $Q$-factor maximum. The q-BICs remain linearly polarized, while the $C$-points appear symmetrically around them, with their separation tunable via $w$. Further increasing $\theta$ shifts both features in momentum space [Fig.~\ref{fig:2}(f)]. At $\theta \approx 71.5^\circ$, each q-BIC merges with a $C$-point, forming a chiral q-BIC.

Fig.~\ref{fig:2}(g) illustrates the distinct behavior of q-BICs and $C$-points near the light cone. While $C$-points (blue/red) follow a trajectory similar to that of accidental BICs in the $w=0$ case (gray line), remaining unaffected by the light cone due to their topological origin~\cite{lepeshov2023topological}, q-BICs (black) are not topologically protected and deviate from linear trajectories as they approach the light cone.

This behavior is governed by the divergence of the radiative local density of states (LDOS). According to Fermi’s golden rule~\cite{ochiai2001nearly}, the radiative $Q$-factor for a mode at wavevector $k_x = 0$, $k_y > 0$ is given by:
\begin{equation}
Q \propto \frac{1}{|A(k_y,k_z)|^2 \rho(k_y,k_z)},
\end{equation}
where $A$ is the mode coupling coefficient \cite{supp}, and $\rho \propto \omega/(c k_z)$ is the 1D vacuum LDOS that diverges near the light cone~\cite{andreani2006photonic}, where $k_z(k_y)=\sqrt{\omega^2/c^2-k_y^2}$ and $\omega$ is the mode frequency. 
This gives:
\begin{equation}
Q(k_y) \propto \frac{k_z(k_y)}{|A(k_y)|^2}.
\label{gamma_rad_final}
\end{equation}

We validate Eq.~\eqref{gamma_rad_final} by calculating $A(k_y)$ as the overlap integral between the incident plane wave and the normalized electric field. As shown in Fig.~\ref{fig:2}(h), $1/|A(k_y)|^2$ (blue line) increases monotonically with $k_y$, while $k_z/|A(k_y)|^2$ (red dashed line) exhibits a clear maximum at $k_y \approx 0.33\cdot2\pi/p$ and decreases rapidly at larger $k_y$. This matches the $Q$-factor from the eigenmode simulations (black line) above the light cone, confirming the mechanism. 
After crossing the light cone, the mode becomes guided and exhibits an infinite $Q$-factor.
We refer to the resulting chiral q-BICs, formed via merging of a q-BIC and a $C$-point near the light cone, as light-cone-proximal chiral q-BICs.

\begin{figure}[htbp]
\centering
\includegraphics[width=1\linewidth]{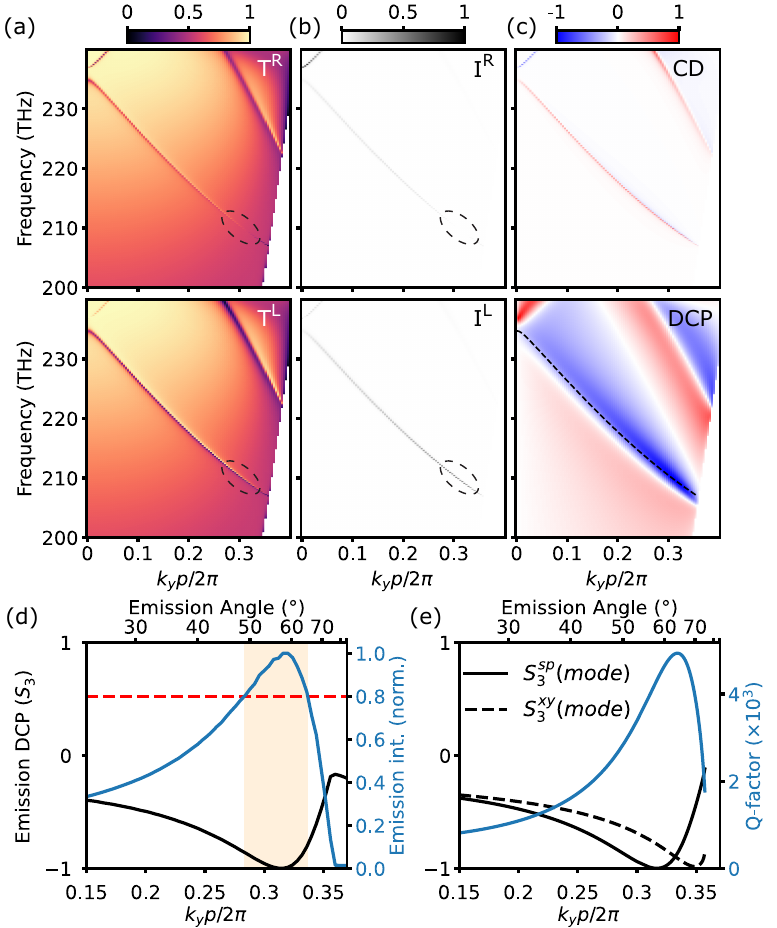}
\caption{
Scattering and emission of light-cone–proximal chiral q-BICs.
(a) Transmission and (b) normalized emission spectra for RCP and LCP light. 
(c) Corresponding circular dichroism (CD) in transmission and degree of circular polarization (DCP) in emission; dashed line indicates resonance dispersion. 
(d) Maximum emission intensity and DCP as a function of Bloch wavevector $k_y$; the shaded region indicates where the emission intensity exceeds $>80\%$.
(e) Comparison of mode 
$Q$--factor (blue) and far-field DCP $S_3$ (black) in 
$sp$ and $xy$ bases (solid vs. dashed lines).
}
\label{fig:3}
\end{figure}

\begin{figure*}
\centering
\includegraphics[width=1\linewidth]{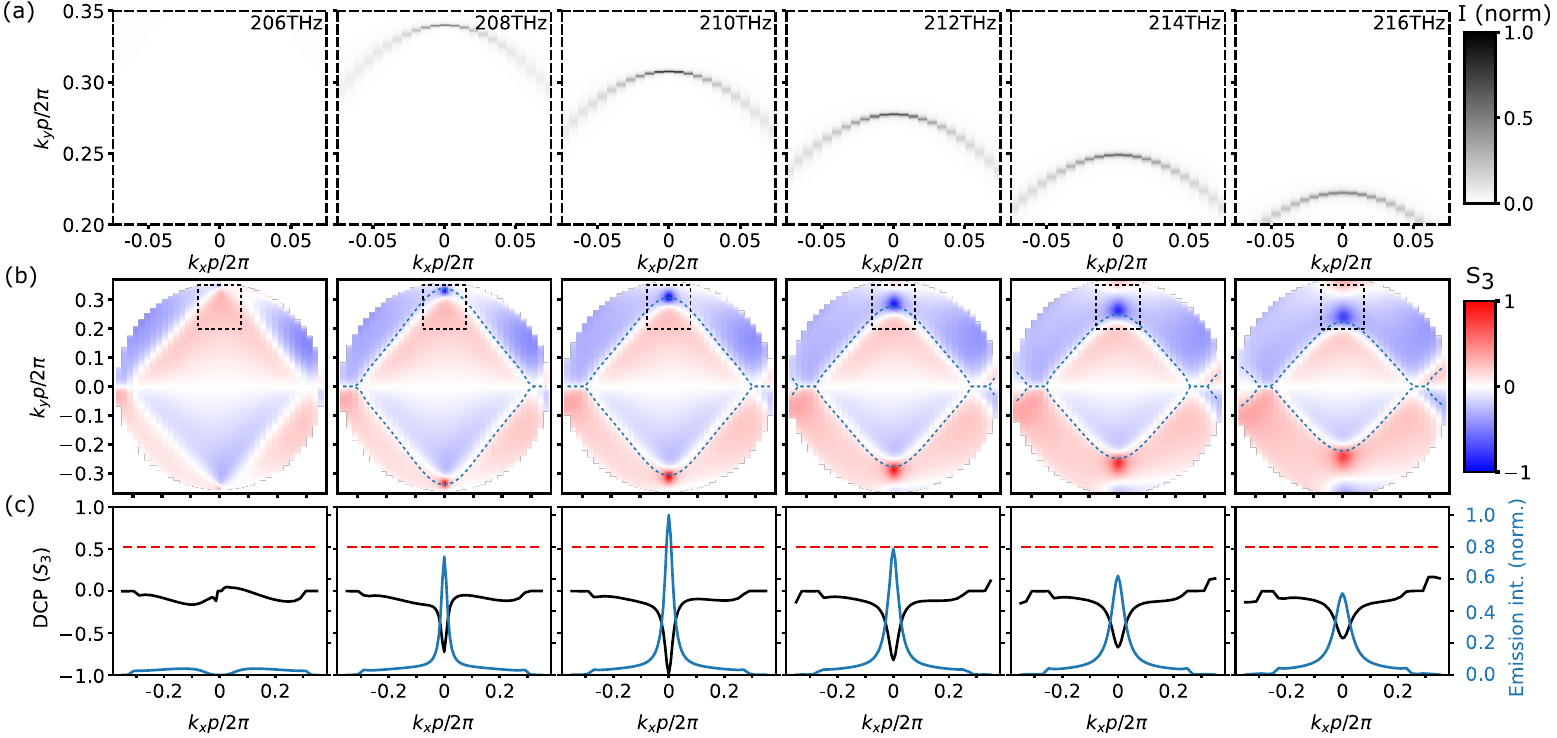}
\caption{Concept of chiral lasing at grazing angles.
(a) Emission enhancement in frequency–momentum space near the chiral q-BIC. 
(b) Corresponding emission DCP maps; dashed squares mark the region shown in (a).
(c) Maximum emission enhancement (blue) vs. tangential wavevector $k_x$ at selected frequencies and corresponding DCP (black). Chiral lasing is expected at 210 THz above the $80\%$ enhancement threshold.
}
\label{fig:4}
\end{figure*}

\textcolor{blue}{\it Emission control for chiral lasing}.--- 
We analyze the scattering and emission properties of the metasurface with $\theta = 72^\circ$ and $w = 40$ nm. The transmission spectra for right- and left-circularly polarized (RCP and LCP) light, $T^R$ and $T^L$, are shown in Fig.~\ref{fig:3}(a). As the resonance approaches the light cone, it becomes increasingly narrow and fully vanishes in $T^R$ at $k_y \approx 0.31 \cdot 2\pi/p$, while remaining visible in $T^L$, indicating the presence of a $C$-point.

A similar asymmetry is observed in the emission enhancement spectra for RCP and LCP components in Fig.~\ref{fig:3}(b), calculated via the reciprocity principle \cite{sauvan2013theory, lobanov2015controlling} (see SI~\cite{supp} for details). The LCP emission dominates at resonance, confirming strong chiral selectivity. The corresponding circular dichroism (CD) in transmission, defined as $\textrm{CD} = (T^R - T^L)/(T^R + T^L)$, and the emission degree of circular polarization (DCP) are shown in Fig.~\ref{fig:3}(c). Notably, the emission DCP provides a clearer picture of the mode’s chirality, as it is less affected by non-resonant contributions.

To further quantify chirality, we evaluate the DCP at the emission enhancement peaks across $k_y$. As shown in Fig.~\ref{fig:3}(d), the peak at $\sim 60^\circ$ corresponds to near-unit chirality ($\textrm{DCP} \approx 1$), and all emission points with enhancement above 80\% maintain $\textrm{DCP} > 0.83$. The momentum-space DCP profile extracted from eigenmode simulations (solid black line) in the $s$–$p$ polarization basis closely matches the emission DCP [Fig.~\ref{fig:3}(e)]. While the $x$–$y$ basis (dashed line) is commonly used for mode characterization, it becomes inaccurate at large angles, necessitating the use of the $s$–$p$ basis.
A slight angular shift is observed between the emission enhancement peak and the $Q$-factor maximum (blue line), attributed to variations in mode volume and near-field parameters.

Finally, we propose a chiral lasing concept based on light-cone-proximal chiral q-BICs. We explore the metasurface emission properties in 3D parameter space $(k_x, k_y, f = \omega/2\pi)$. Figs.~\ref{fig:4}(a,b) show the emission intensity enhancement and DCP at selected frequencies from 206 THz to 216 THz near the q-BIC resonance. Maximum emission occurs at $(k_x, k_y, f) = (0, 0.31 \cdot 2\pi/p, 210~\text{THz})$ and rapidly decreases with variation in any parameter. This is confirmed by the plots of peak enhancement versus  $k_x$ [Fig.~\ref{fig:4}(c)]. Assuming lasing occurs above an 80\% enhancement threshold, these results highlight the metasurface as a promising platform for highly directional, circularly polarized emission at large angles.

\textcolor{blue}{\it Conclusion \& Discussion}.---
We introduce a novel mechanism for realizing light-cone-proximal chiral q-BICs at large oblique angles, enabled by the divergence of the radiative local density of states near the light cone. By breaking in-plane mirror symmetries within a monoclinic lattice and tuning its lattice angle, we achieve robust chiral q-BICs in fully planar dielectric metasurfaces that are compatible with standard fabrication processes on low-index substrates.

A key advantage of this design is the tunability of the q-BICs’ position in momentum space via the lattice angle. This enables efficient control over the resonant emission direction across a broad range of polar angles ($0^\circ$ to $70^\circ$). At a specific lattice angle, the system exhibits near-unity circular dichroism and generates circularly polarized far-field emission, all while bypassing the normal-incidence constraint typically associated with conventional chiral q-BICs.
These resonances combine high-$Q$ factors and excellent polarization control comparable with those observed at the $\Gamma$-point, yet are accessible at large emission angles. This makes light-cone-proximal chiral q-BICs highly promising for directional chiral lasing, potentially supporting dual beams propagating at approximately $\sim60^\circ$  from the surface normal.

Moreover, the planar design avoids the need for out-of-plane symmetry elements, offering greater structural flexibility and ease of fabrication. Suitable platforms for experimental realization include silicon metasurfaces integrated with quantum emitters or multiple-quantum-well semiconductor metasurfaces.

Beyond dual-beam operation, our approach can be extended to support multiple light-cone-proximal q-BICs at distinct momentum points, enabling selective suppression or generation of multibeam emission with different polarizations. Overall, the concept of light-cone-proximal q-BICs opens new avenues for the development of compact, efficient devices that generate directional, circularly polarized light at large angles, with impactful applications in polarimetric sensing, integrated optical computing, nonlinear optics, and quantum photonics.

\begin{acknowledgments}
This work was supported by the National Research Foundation, Singapore (NRF-CRP26-2021-0004, NRF-CRP31-0007), the Ministry of Education, Singapore (MOE-T2EP50223-0001), the Agency for Science, Technology and Research, Singapore (MTC IRG M24N7c0083), and the Singapore University of Technology and Design (Kickstarter Initiative SKI 2021-04-12). D.G. acknowledges support from the SUTD-NUS Ph.D. Research Scholarship Scheme. K.K. acknowledges funding from the Australian Research Council (DECRA Fellowship DE250100419). C.-W.Q. acknowledges financial support from the Ministry of Education, Singapore (A-8002152-00-00, A-8002458-00-00), and the National Research Foundation, Singapore (Competitive Research Programme Awards NRF-CRP26-2021-0004, NRF-CRP30-2023-0003).
\end{acknowledgments}

\bibliographystyle{elsarticle-num}


\end{document}


\title{Supplemental Material for ``Cone-proximity quasi-BIC resonances for chiral lasing at grazing angles''}

\author{Dmitrii Gromyko}
\affiliation{Science, Mathematics, and Technology (SMT), Singapore University of Technology and Design (SUTD), 8 Somapah Road, Singapore 487372.}
\affiliation{Department of Electrical and Computer Engineering, National University of Singapore, 4 Engineering Drive 3, Singapore 117583.}

\author{Cheng-Wei Qiu}
\email{chengwei.qiu@nus.edu.sg}
\affiliation{Department of Electrical and Computer Engineering, National University of Singapore, 4 Engineering Drive 3, Singapore 117583.}

\author{Kirill Koshelev}
\email{kirill.koshelev@anu.edu.au}
\affiliation{Department of Electronic Materials Engineering, Research School of Physics,
Australian National University, Canberra, ACT, Australia}

\author{Lin Wu}
\email{lin\_wu@sutd.edu.sg}
\affiliation{Science, Mathematics, and Technology (SMT), Singapore University of Technology and Design (SUTD), 8 Somapah Road, Singapore 487372.}

\maketitle
\tableofcontents
\newpage
\section{Numerical methods}
\subsection{RCWA simulations}
The results presented in the main text were simulated using the rigorous coupled-wave analysis (RCWA) open-access simulation package Inktone\cite{alexysong_inkstone}. RCWA, also known as the Fourier modal method \cite{whittaker1999scattering, tikhodeev2002quasiguided}, is perfectly suited for structures that can be divided into layers that are strictly periodic along the lateral directions ($xy$-plane) and homogeneous along the vertical $z$-direction. Due to spatial periodicity, all material properties, such as permittivity $\varepsilon$ and permeability $\mu$, and electromagnetic fields $\mathbf{F}(\mathbf{r})\equiv\mathbf{E}(\mathbf{r}),\mathbf{H}(\mathbf{r})$ inside each layer can be presented in the form of a Fourier series with a finite number of harmonic terms:
\begin{equation}
\varepsilon(\mathbf{r}_{||})=\sum_\mathbf{G}\varepsilon_\mathbf{G}\exp(i\mathbf{G}\mathbf{r}_{||}); \quad \mathbf{F}(\mathbf{r}_{||},z)=\sum_\mathbf{G}\mathbf{F}_\mathbf{G}(z)\exp(i(\mathbf{k}_{||}+\mathbf{G})\mathbf{r}_{||}), 
\end{equation}
where $\mathbf{r}_{||}=(x,y)$, $\mathbf{k}_{||}$ is the in-plane Bloch vector, $\mathbf{G}=n_1\mathbf{G}_1+n_2\mathbf{G}_2$, $\mathbf{G}_{1,2}$ are the basis vectors of the reciprocal lattice, $n_{1,2}$ take any integer values from $-N_{1,2}$ to $N_{1,2}$.
In this case, Maxwell's equations within each metamaterial layer can be reduced to a first-order differential equation system
\begin{equation}
    -i\partial_z\mathcal{F}_{||}=\mathcal{M}(\varepsilon,\mu)\mathcal{F}_{||},
\end{equation}
which is used to identify eigenmodes $\mathcal{F}^{(i)}_{||}$ and corresponding propagation constants $k_{z,i}$. Vectors $\mathcal{F}^{(i)}_{||}$ that represent the modes in each layer are composed of the Fourier components of the tangential component of the electromagnetic fields and propagate along the vertical direction as $\exp(\pm ik_{z,i} z)$.  Modes of adjacent layers are stitched together at the boundary between the layers according to Maxwell's boundary conditions, which allows for evaluating the electric fields in the entire structure iteratively, layer by layer. 

This approach is efficiently combined with the scattering matrix (S-matrix) formalism, which directly relates the amplitudes of the incoming and outgoing modes of a given layer or the whole structure:
\begin{equation}
    \mathcal{A}^{out}=S\mathcal{A}^{in}.
\end{equation}
If the structure of interest is sandwiched between two homogeneous semi-infinite half-spaces with permittivities $\varepsilon_a$ ($a=1,2$, for top and bottom half-spaces), the modes of the outer layers are simply plane waves with wavevectors 
\begin{equation}
    \mathbf{k}^{\pm}_{\mathbf{G},a}=(k_{x,\mathbf{G}},k_{y,\mathbf{G}}, \pm k_{z,\mathbf{G},a}),
\end{equation}
where 
\begin{equation}
    k_{x,\mathbf{G}}=k_x+G_x,\quad k_{y,\mathbf{G}}=k_y+G_y, \quad k_{z,\mathbf{G},a}=\sqrt{\omega^2\varepsilon_a/c^2-k_{x,\mathbf{G}}^2-k_{y,\mathbf{G}}^2}.
\end{equation}

Once the full S-matrix of the structure is evaluated, it can be reduced to the truncated S-matrix that only includes scattering in the radiative channels of the structure, i.e., those that can be accessed in the far-field measurements and feature $\mathrm{Im}\,k_{z,\mathbf{G},a}=0$. In the case when only the main diffraction channel is open for radiation, the S-matrix has a simple 4-by-4 form:
\begin{equation}
    \begin{pmatrix}
        u_s^1\\
        u_p^1\\
        d_s^2\\
        d_p^2
    \end{pmatrix}=\begin{pmatrix}
        r_{ss}& r_{sp}& t'_{ss}& t'_{sp}\\
        r_{ps}& r_{pp}& t'_{ps}& t'_{pp}\\
        t_{ss}& t_{sp}& r'_{ss}& r'_{sp}\\
        t_{ps}& t_{pp}& r'_{ps}& r'_{pp}
    \end{pmatrix}\begin{pmatrix}
        d_s^1\\
        d_p^1\\
        u_s^2\\
        u_p^2
    \end{pmatrix},
\end{equation}
where $d$ and $u$ denote amplitudes of the waves propagating downwards and upwards, subscripts $s, p$ denote the polarization of waves, $r$ and $t$ are complex amplitude transmission and reflection coefficients of the structure, and indices $1,2$ indicate waves in the top and bottom half-spaces (superstrate and substrate) respectively. Due to the direct consideration of multiple radiative and nonradiative (evanescent) channels with explicitly calculated $k_{z,\mathbf{G},a}$, RCWA methods are considered much more accurate near the light cones and diffraction thresholds manifesting as $k_{z,\mathbf{G},a}=0$. 

Besides transmission and reflection coefficients, one can also evaluate the resonant modes of a layered structure, which are represented by first-order poles of the S-matrix:
\begin{equation}
    S(\omega)=S_\text{bg}(\omega)+\sum_n\dfrac{\left|O_n\right\rangle\left\langle I_n\right|}{\omega-\omega_n},
\end{equation}
where $S_\text{bg}(\omega)\approx\text{const}$ is slowly-varying nonresonant term, $\omega_n$ are complex resonant frequencies of the eigenmodes and $\left|O_n\right\rangle$, $\left\langle I_n\right|$ are the output and input resonant vectors defining coupling of the resonant mode $n$ with the modes of the outer layers. To find the resonant modes of the structure of interest, we identified the poles using Newton's root-finding method for the inverse S-matrix on the complex frequency plane \cite{gippius2010resonant}. The electric fields of the modes were evaluated using the resonant output vectors $\left|O_n\right\rangle$ as outgoing amplitudes and zero incoming amplitudes. 

In our simulations, we used $N_{1,2}=N=7$ with a total number of $(2N+1)^2$ harmonics. All parameters of the metasurface are indicated in the main text. 

For the metasurfaces shown in Fig. 1 of the main text, we used the following parameters. Fig. 1(b): metasurface with square lattice with period $p=600$ nm, thickness $h=100$ nm, triangular holes with side length $l=450$ nm, permittivity $\varepsilon=12.25$, surrounded by air, the resonant mode near 240 THz. Fig. 1(c): metasurface with square lattice with period $p=340$ nm, thickness $h=220$ nm, trapezoidal holes with side length $w=210$ nm, perturbation angle $\alpha=0.12$ rad, vertical inclination angle $\phi=0.1$ rad (see the original Ref. \cite{chen2023observation}), permittivity $\varepsilon=2.13^2$, surrounded by silica with $n=1.46$.

\subsection{Reciprocity principle for emission simulations}
Emission from a periodic distribution of emitters with defined orientation coupled to a periodic resonant environment, such as a metasurface, can be calculated using the RCWA and S-matrix formalism directly \cite{lobanov2012emission}. Alternatively, it is possible to simulate emission and scattering properties in the case when emitters are defined as point objects with a certain polarizability tensor \cite{fradkin2019fourier}. However, in the case of randomly oriented dipoles located in the near-field of the metasurface, it is more convenient to use the Lorentz reciprocity theorem for the evaluation of the emission signal intensity \cite{sauvan2013theory, lobanov2015controlling}. 
 
 First, consider a periodic dipole density $\mathbf{P}(\mathbf{r})$ oscillating at frequency $\omega$ in a spatially periodic structure surrounded by a medium with refractive index $n$. The reciprocity theorem states that the intensity of the signal with polarization $\sigma$ emitted by these dipoles in a given direction defined by angles $\Theta$ and $\Phi$ is proportional to the squared integral of this dipole density and the electric field induced at the location of the dipoles by a hypothetical plane wave with polarization $\sigma$ incident on the structure from the same direction $(\Theta,\Phi)$:
\begin{equation}
    I_\sigma(\Theta,\Phi)\propto \left|\int_{V_\mathrm{UC}}\tilde{\mathbf{E}}(\mathbf{r})\cdot\mathbf{P}(\mathbf{r})d^3\mathbf{r}\right|^2,
    \label{Lorentz}
\end{equation}
where $\tilde{\mathbf{E}}_\sigma(\mathbf{r})$ is the electric field induced in the metasurface by a plane wave $\mathbf{E}_\sigma \exp(i\omega t+i\mathbf{kr})$, $\mathbf{k}=\dfrac{n\omega}{c}\{\sin\Theta\cos\Phi, \sin\Theta\sin\Phi, \cos\Theta\}$. The integral is evaluated over the unit cell volume $V_\mathrm{UC}$. When the emitters exhibit random orientation in space and are uniformly distributed in a certain area, one should average the local field intensity over all possible orientations, which leads to a simple formula:
\begin{equation}
    \langle I_\sigma(\Theta,\Phi)\rangle\propto \int_{\mathbf{r}:\; \mathrm{P}(\mathbf{r})\ne0}\tilde{I}_\sigma(\mathbf{r})d^3\mathbf{r},
    \label{unpolarized_itensity}
\end{equation}
where $\tilde{I}_\sigma(\mathbf{r})=|\tilde{\mathbf{E}}_\sigma(\mathbf{r})|^2$ is the local field intensity.

To evaluate the enhancement of circularly polarized emission from the metasurface, we take an incident wave with a unitary amplitude and $\sigma=\text{RCP,LCP}$. We note that without the structure, a periodic arrangement of randomly polarized uncorrelated dipoles radiates isotropically, which agrees with $\tilde{I}_\sigma(\mathbf{r})=\text{const}$ in free space regardless of the incidence direction. The emission enhancement can be calculated as a ratio of the intensities calculated according to Eq. \eqref{unpolarized_itensity} with and without the metasurface. The results of these simulations are shown in Figs. 3 and 4 of the main text.

\subsection{COMSOL simulations}
The resonant modes of the metasurface and their properties were found using an Eigenfrequency study in COMSOL. The simulation domain comprised one unit cell of the metasurface (monoclinic lattice shown in the main text, Fig. 2(a)), sandwiched between two air domains of $6000$ nm thickness. Periodic conditions with Floquet periodicity were used on the lateral boundaries with the Bloch wavevector components $k_x, k_y$ specified explicitly. $3000$ nm thick PMLs were placed on the upper and lower boundaries of the simulation domain. We used a manually defined mesh with a maximum element growth rate of 1.3, a curvature factor of 0.2, and a resolution of narrow regions factor of 1. The maximum element size of the mesh in the air regions was 200 nm (denser near the material boundaries) and 40 nm within the metasurface unit cell (denser near the material boundaries, about 10 nm in lateral dimensions in the rectangular hole). The maximum element size in PMLs was 100 nm with explicitly defined mesh distribution along the vertical direction (30 elements per PML domain). Meshes on the opposite lateral boundaries of the simulation domain were copies. Such domain thickness and mesh density were chosen to minimize backward reflection of the PMLs at large emission angles.

The eigenfrequencies search was conducted near 208 THz, with the desired number of eigenfrequencies equal to 7. The eigenfrequencies were sorted and separated based on the requirement of continuity of the real and imaginary parts of the resonant frequency to extract the mode of interest. Since COMSOL uses time dependence $(i\omega t)$, the far-field polarization of a wave $\mathbf{E}^\text{ff}e^{-i(k_xx+k_yy\pm k_zz)}$ was extracted from the averaged electric fields calculated on the top and bottom boundaries between air domains and PMLs $A^{t,b}$ as
\begin{equation}
    \mathbf{E}^\text{ff}=\dfrac{1}{A_{UC}}\int_{A^{t,b}} \mathbf{E}(\mathbf{r})\exp(ik_xx+ik_yy)dA,
\end{equation}
where $A_{UC}$ is the area of the unit cell, equal to the area of boundaries $A^{t,b}$.

\section{Controlling the radiative Q-factor of cone-proximity chiral qBICs using structural asymmetry}
\begin{figure}[h!]
    \centering
    \includegraphics[width=1\linewidth]{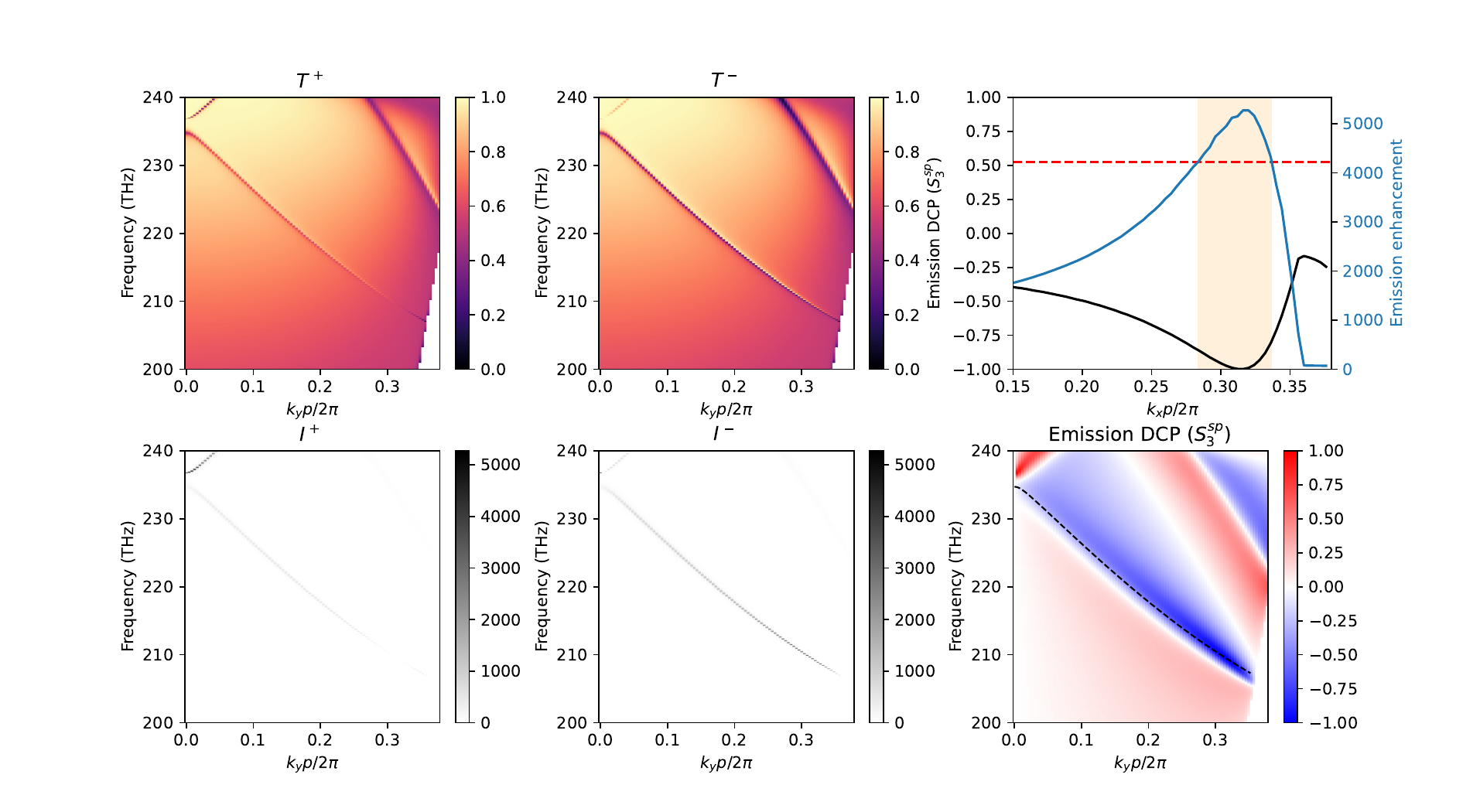}
    \caption{Scattering and emission properties of cone-proximity chiral q-BICs in a metasurface with $\theta=72^\circ$, $w=40$ nm.}
    \label{fig:S1_th=72_w=40}
\end{figure}

\begin{figure}[h!]
    \centering
    \includegraphics[width=1\linewidth]{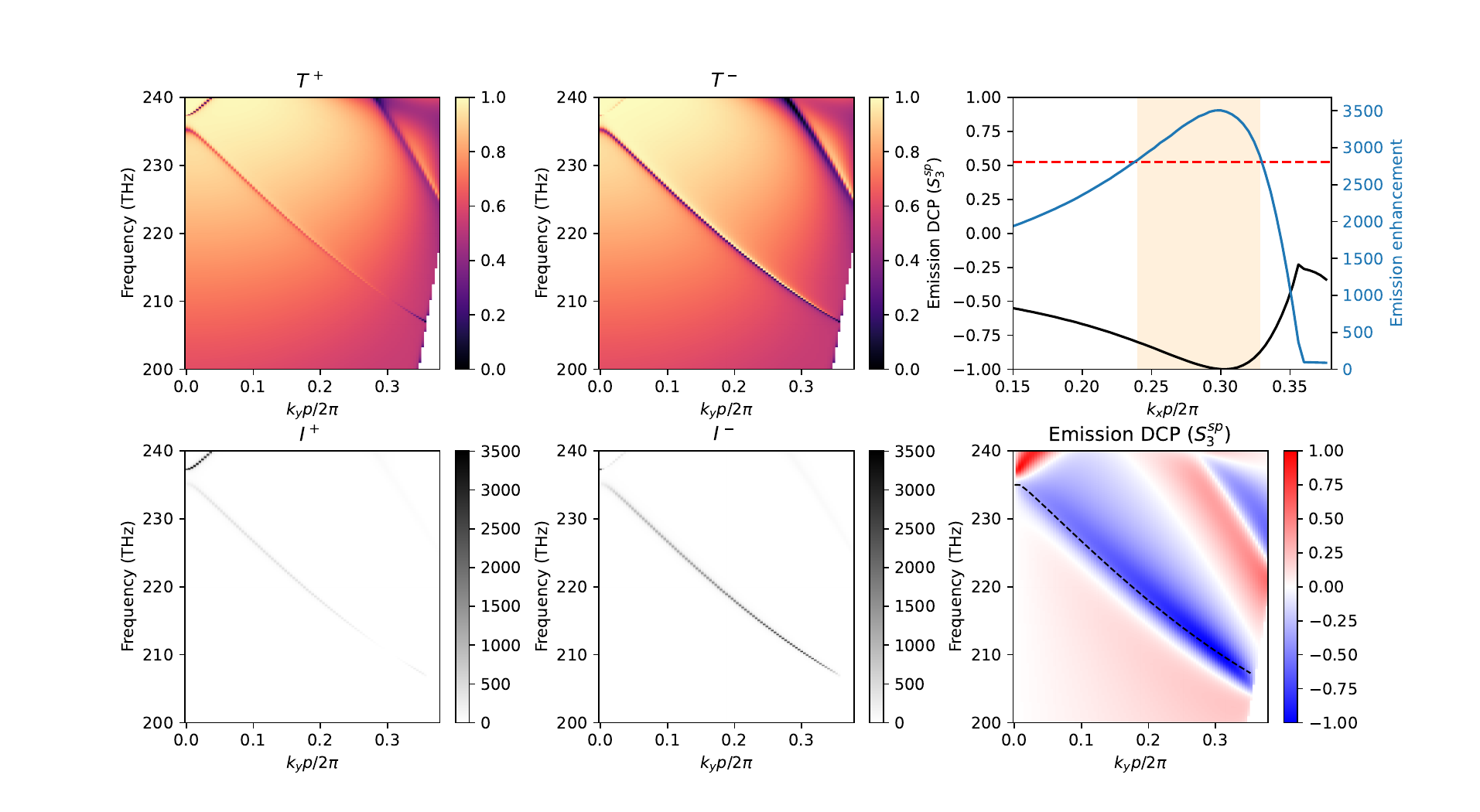}
    \caption{Scattering and emission properties of cone-proximity chiral q-BICs in a metasurface with $\theta=73^\circ$, $w=50$ nm.}
    \label{fig:S1_th=73_w=50}
\end{figure}

\begin{figure}[h!]
    \centering
    \includegraphics[width=1\linewidth]{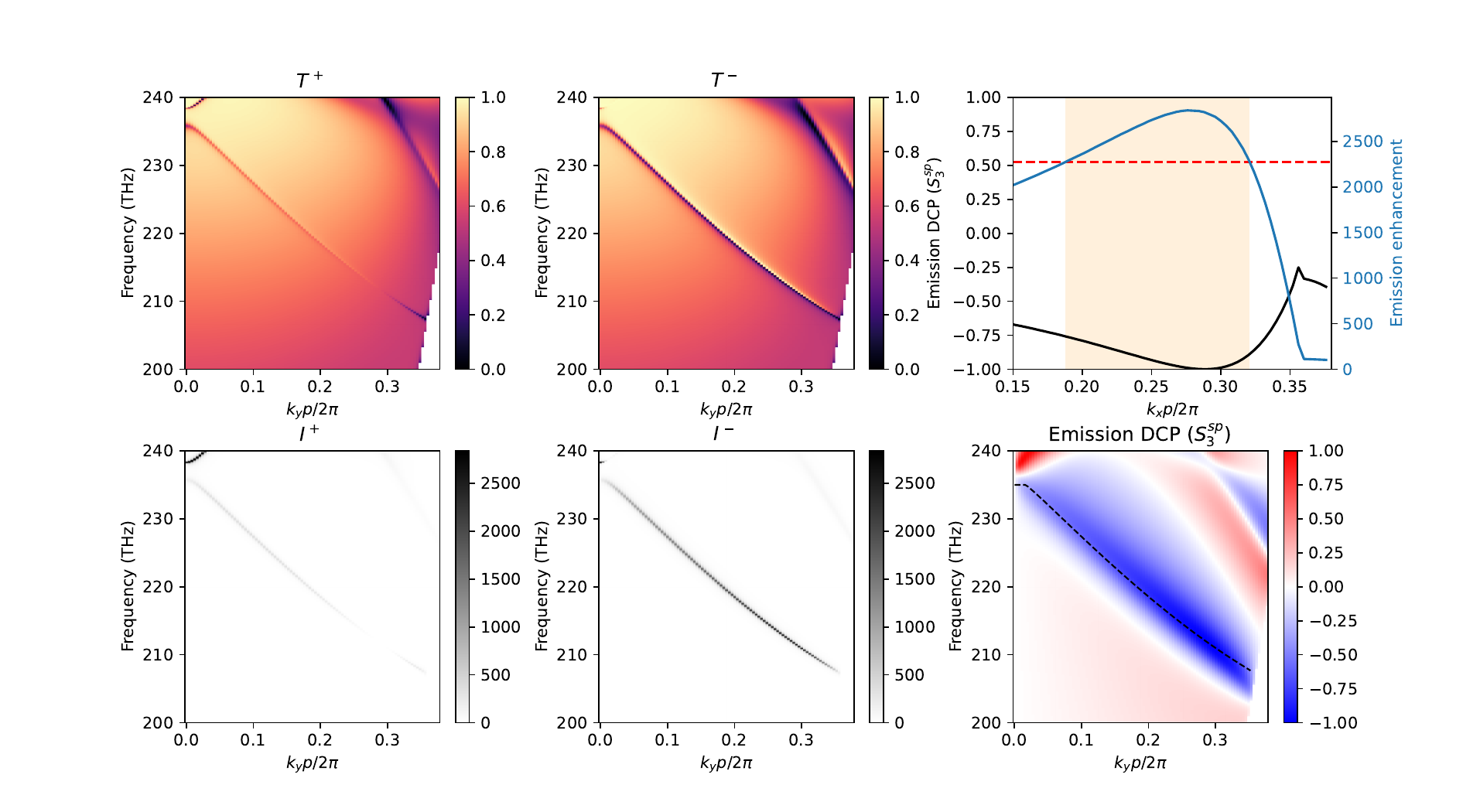}
    \caption{Scattering and emission properties of cone-proximity chiral q-BICs in a metasurface with $\theta=74^\circ$, $w=60$ nm.}
    \label{fig:S1_th=74_w=60}
\end{figure}

One of the most important properties of qBICs is the ability to manipulate their Q-factor by means of structural asymmetry \cite{koshelev2018asymmetric}. Here, we demonstrate that cone-proximity qBICs also allow for efficient Q-factor manipulation using the size of the rectangular hole that we use as a symmetry perturbation. Figs. \ref{fig:S1_th=72_w=40}-\ref{fig:S1_th=74_w=60} demonstrate the scattering and emission spectra of metasurfaces with the rectangular hole sizes $w=40-60$ nm. Variation of $w$ results in a significant modulation of the maximum Q-factor from $\sim 5000$ to $\sim 1000$, which can be used to achieve maximum resonant enhancement at a given level of nonradiative losses, known as critical coupling. At the same time, high emission DCP is achieved in all metasurface configurations. 

\newpage
\section{RCWA and COMSOL simulations comparison}
\begin{figure}[h!]
    \centering
    \includegraphics[width=0.8\linewidth]{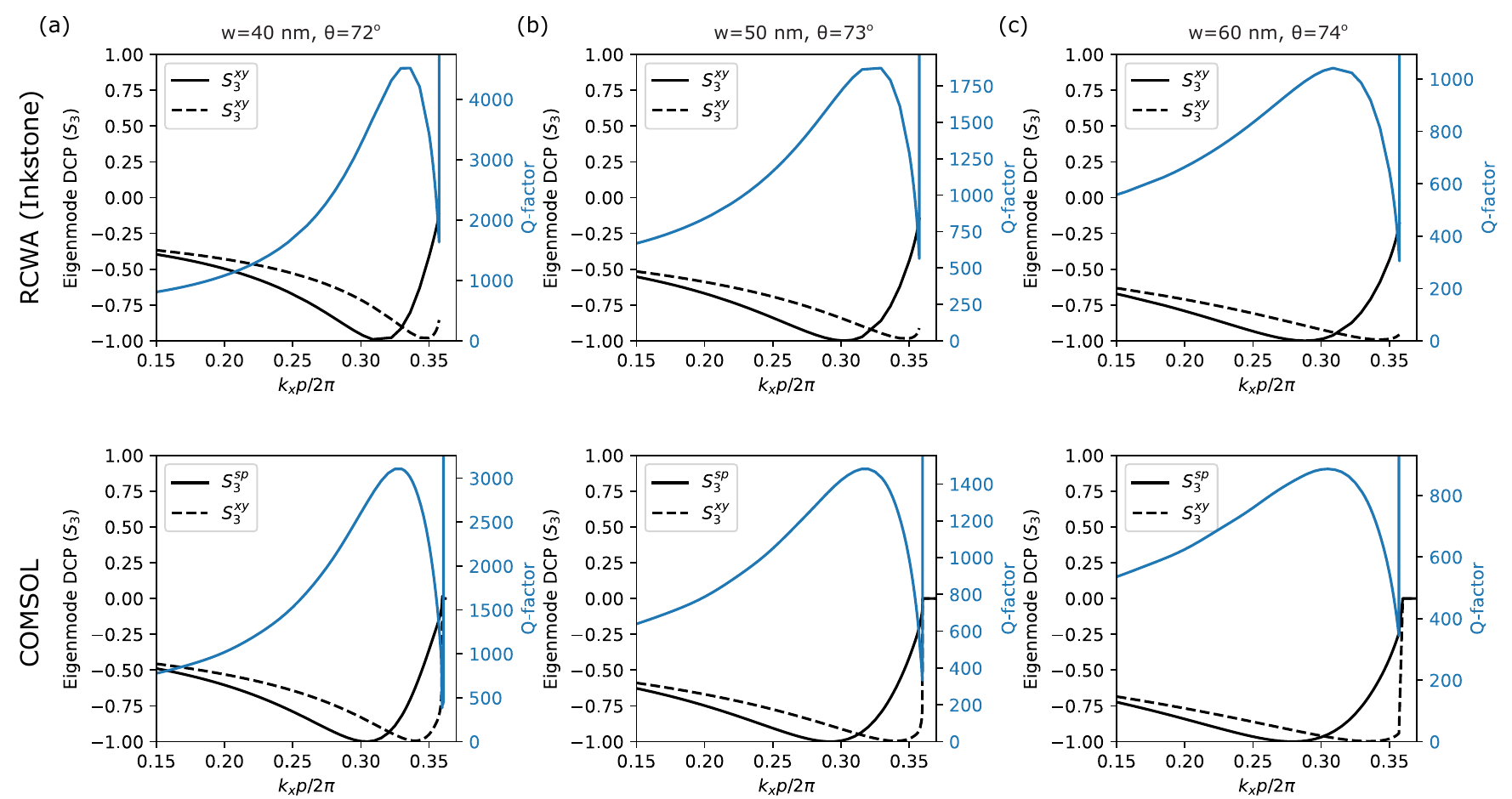}
    \caption{RCWA (Inkstone) and FEM (COMSOL) eigenmode simulations comparison for metasurfaces exhibiting chiral cone-proximity qBIC. The parameters of metasurfaces are: (a) $w=40$ nm, $\theta=72^\circ$, (b) $w=50$ nm, $\theta=73^\circ$, (c) $w=60$ nm, $\theta=74^\circ$.  }
    \label{COMSOL_vs_RCWA}
\end{figure}

We compare the results extracted from RCWA and COMSOL eigenmode simulations and find good agreement between the DCP values, position of the Q-factor maximum, and the overall shape of $Q(k_y)$ dependence, see Fig. \ref{COMSOL_vs_RCWA}. The moderate discrepancy in the absolute Q-factor values calculated using COMSOL and the RCWA tool can be attributed to the fundamental difference between these two numerical methods, which nonetheless yield similar results.
\section{Resonant coefficient of far-field coupling and radiative losses}
The resonant coefficient of far-field coupling $A$ can be evaluated from the  reciprocity as the overlap integral of the incident plane wave $\mathbf{E}_\mathrm{inc}\exp(i\mathbf{kr})$ at frequency $\omega=\text{Re}(\omega_r)$ and the normalized electric field $\mathbf{E}_\mathrm{res}(k_x,k_y;\mathbf{r})$ of the resonant mode with frequency $\omega_r$:
\begin{equation}
A(k_x,k_y)\propto\int_{V_\text{UC}}\hspace{-0.4cm}[\varepsilon(\mathbf{r})-1]\mathbf{E}_\mathrm{res}(k_x,k_y;\mathbf{r})\cdot\mathbf{E}_\mathrm{inc}\exp(i\mathbf{kr}){\rm d}^3\mathbf{r}.
\label{eq_overlap}
\end{equation}
where $\varepsilon(\mathbf{r})-1$ is the difference between the permittivities of the metasurface and the surrounding material (air in our case). The electric field $\mathbf{E}_\mathrm{res}$ is normalized according to \cite{weiss2018calculate}:
\begin{equation}
    \int\mathbf{E}_\mathrm{res}(k_x,k_y;\mathbf{r})\cdot \mathbf{E}^*_\mathrm{res}(-k_x,-k_y;\mathbf{r})\varepsilon(\mathbf{r}){\rm d}^3\mathbf{r}=1,
\end{equation}
where the integral is taken within one unit cell along the lateral dimensions and in a sufficiently large area along the vertical $z$-direction (we take $z\in [-5h,6h]$). Although rigorously, this integral does not converge in the whole space unless the mode is nonradiative, this is a valid approximation for high-Q modes. A more general and rigorous approach for normalization of highly radiative modes that also includes integrals on the top and bottom boundaries of the simulation domain can be found in Refs. \cite{weiss2018calculate}.

 The radiative losses of a resonance can be represented as a product of the squared coupling coefficient $|A(k_x,k_y)|^2$, which shows how well the mode can be excited by an incident plane wave, and an LDOS $\rho(\omega,k_x,k_y)$ which is the number of states (available plane waves with a given tangential wavevector $(k_x,k_y)$) per an infinitesimal range of frequencies $d\omega$:
\begin{equation}
\gamma_r(k_x,k_y)\propto|A(k_x,k_y)|^2\cdot\rho(\omega,k_x,k_y).
\end{equation}
Consequently, we see that the radiative losses diverge as the resonance approaches the light cone and the quality factor $Q=\omega/2\gamma_r$ rapidly decreases, as can be seen in Fig. \ref{COMSOL_vs_RCWA}.
\section{Impact of the substrate}
\begin{figure}[h!]
    \centering
    \includegraphics[width=1\linewidth]{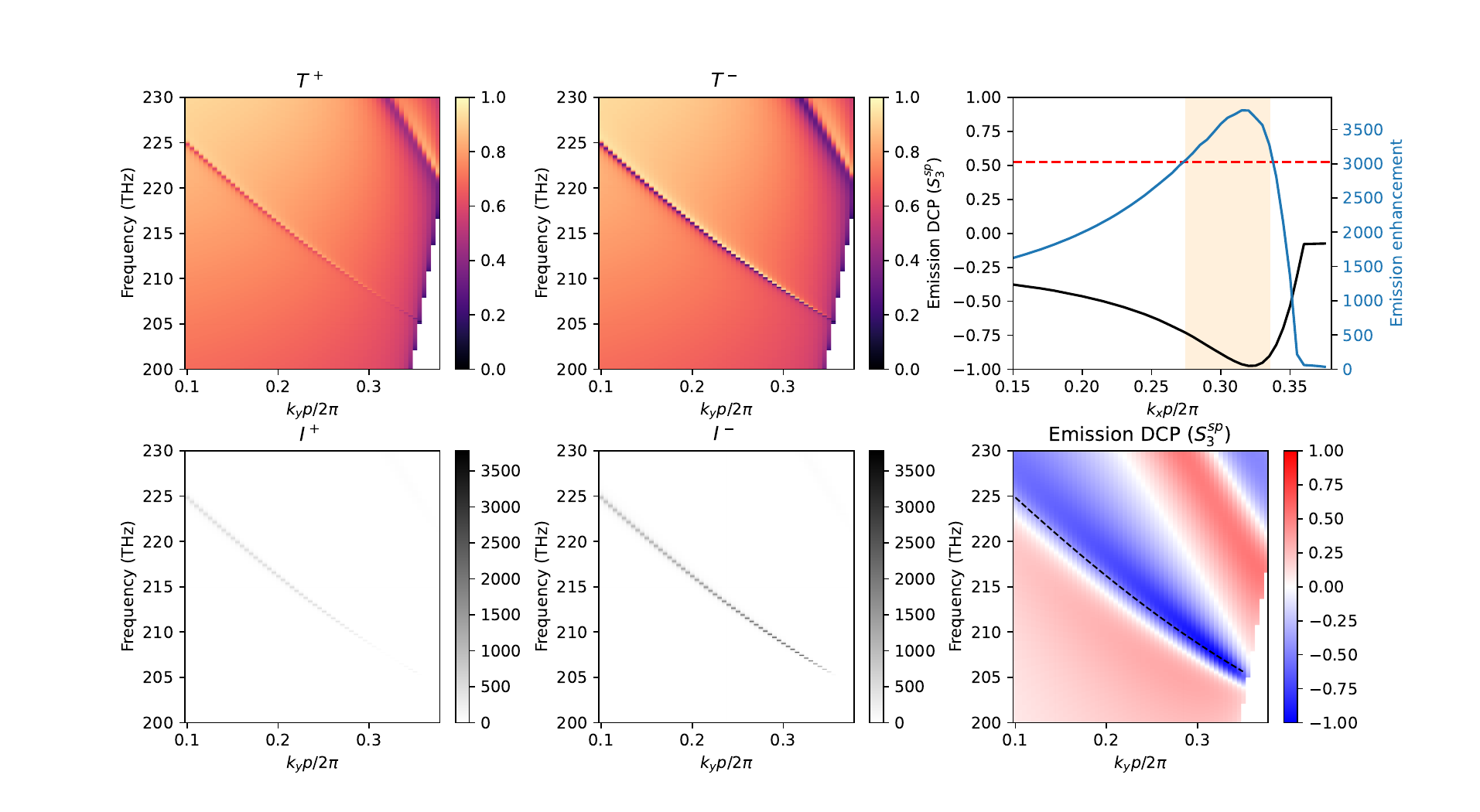}
    \caption{Scattering and emission properties of cone-proximity chiral q-BICs in a metasurface with $\theta=72^\circ$, $w=40$ nm on a SiO$_2$ substrate ($n=1.45$).}
    \label{Spectra_on_Sub}
\end{figure}

Modern nanofabrication technology enables the fast and convenient production of single-layer metasurfaces on insulator substrates with low refractive indices, such as SiO$_2$. While the main text considers an ideal suspended structure, it is essential to verify the formation mechanism of cone-proximity qBICs in structures with a substrate. The scattering and emission properties of the metasurface with $w=40$ nm and $\theta=72^\circ$ placed on a substrate with refractive index $n=1.45$ are demonstrated in Fig. \ref{Spectra_on_Sub}. One can see that the presence of the substrate does not significantly affect the position of the maximum emission enhancement, the degree of circular polarization, or the dispersion of modes.  

\bibliographystyle{elsarticle-num}
\bibliography{SPDC_lib}